\documentclass[letterpaper,twocolumn,10pt]{article}
\usepackage{usenix2019_v3}

\usepackage{tikz}
\usepackage{amsmath}
\usepackage{listings}
\usepackage{subfigure}
\usepackage{multirow}
\usepackage{booktabs}
\usepackage{colortbl}
\usepackage{bbding}
\usepackage{pifont}
\usepackage{color}

\usepackage{xcolor}
\usepackage[linesnumbered,ruled,vlined]{algorithm2e}
\SetCommentSty{mycommfont}

\SetKwInput{KwInput}{Input}                % Set the Input
\SetKwInput{KwOutput}{Output}              % set the Output
\usepackage{algpseudocode}
\algtext*{EndWhile}% Remove "end while" text

\usepackage{diagbox}
\usepackage{booktabs}

\begin{document}
%-------------------------------------------------------------------------------

% codes
\definecolor{mygreen}{rgb}{0,0.6,0}
\definecolor{mygray}{rgb}{0.5,0.5,0.5}
\definecolor{mymauve}{rgb}{0.58,0,0.82}
\lstset{ %
backgroundcolor=\color{white},   % choose the background color
basicstyle=\footnotesize\ttfamily,        % size of fonts used for the code
columns=fullflexible,
breaklines=true,                 % automatic line breaking only at whitespace
captionpos=b,                    % sets the caption-position to bottom
tabsize=4,
commentstyle=\color{mygreen},    % comment style
escapeinside={\%*}{*)},          % if you want to add LaTeX within your code
keywordstyle=\color{blue},       % keyword style
stringstyle=\color{mymauve}\ttfamily,     % string literal style
frame=single,
rulesepcolor=\color{red!20!green!20!blue!20},
% identifierstyle=\color{red},
language=c++,
}

%don't want date printed
% \date{}

% \title{IDT: Transparent In-band Distributed Tracing in Kernel}
\title{Nahida: In-Band Distributed Tracing with eBPF}

% for single author (just remove % characters)
\author{
{\rm Wanqi Yang}\\
Sun Yat-sen University \\
yangwq23@mail2.sysu.edu.cn
\and
{\rm Pengfei Chen}\\
Sun Yat-sen University \\
chenpf7@mail.sysu.edu.cn
\and
{\rm Kai Liu}\\
Alibaba Group \\
qianlu.lk@alibaba-inc.com
\and
{\rm Huxing Zhang}\\
Alibaba Group \\ 
huxing.zhx@alibaba-inc.com
}
% end author

\maketitle

%-------------------------------------------------------------------------------
\begin{abstract}
%-------------------------------------------------------------------------------
% Your abstract text goes here. Just a few facts. Whet our appetites.
% Not more than 200 words, if possible, and preferably closer to 150.

Microservices are commonly used in modern cloud-native applications to achieve agility. However, the complexity of service dependencies in large-scale microservices systems can lead to anomaly propagation, making fault troubleshooting a challenge. To address this issue, distributed tracing systems have been proposed to trace complete request execution paths, enabling developers to troubleshoot anomalous services. However, existing distributed tracing systems have limitations such as invasive instrumentation, trace loss, or inaccurate trace correlation. To overcome these limitations, we propose a new tracing system based on eBPF (extended Berkeley Packet Filter), named Nahida, that can track complete requests in the kernel without intrusion, regardless of programming language or implementation. Our evaluation results show that Nahida can track 
over 92\% of requests with stable accuracy, even under the high concurrency of user requests, while the state-of-the-art non-invasive approaches can not track any of the requests. Importantly, Nahida can track requests served by a multi-threaded application that none of the existing invasive tracing systems can handle by instrumenting tracing codes into libraries. Moreover, the overhead introduced by Nahida is negligible, increasing service latency by only 1.55\%--2.1\%.
Overall, Nahida provides an effective and non-invasive solution for distributed tracing.

\end{abstract}

%-------------------------------------------------------------------------------
\section{Introduction}
%-------------------------------------------------------------------------------
% todo: cite
% 1. Background
% 2. Problem
Modern large-scale distributed systems are often implemented in microservices architecture due to high scalability and reliability~\cite{coulouris2005distributed, mullender1990distributed, van2016brief, waldo2005note, microservice}.
However, managing such systems can be challenging, especially when they fail. The complexity of service dependencies and anomaly propagation in such systems make troubleshooting difficult. To address this issue, distributed tracing systems~\cite{pinpoint, xtrace, dapper, zipkin, jaeger, opentelemetry, opentracing, fay, skywalking} have been proposed to help developers and operators troubleshoot anomalous services, and are now widely used in the production environment.

Distributed tracing systems track the end-to-end requests, recording calls from downstream services to upstream services  along with their duration. 
With tracing data, developers and operators can easily troubleshoot anomalous services, even when anomaly propagation occurs, thereby facilitating system failure resolution. 
However, tracing end-to-end requests in a large-scale distributed system can be difficult, as in such systems, an end-to-end request often spans multiple services across different physical machines and geographical locations, where services may be implemented in different high-level languages and communicate in different protocols. 

% 3. Existing solutions & Limitation
Distributed tracing systems are mainly chosen to be implemented in an invasive manner to meet the challenges above, such as Zipkin~\cite{zipkin}, Jaeger~\cite{jaeger}, OpenTelemetry~\cite{opentelemetry}, OpenTracing~\cite{opentracing}, X-Trace~\cite{xtrace} and Dapper~\cite{dapper}. 
These invasive tracing systems require developers to weave tracing code into application source code or dependent third-party libraries, thus automatically generating traces as the application runs.
However, these invasive tracing systems have their downside including requiring explicit and manual tracing code instrumentation and lacking support for hot updates. The frequent updates to applications and dependent third-party libraries require developers to spend significant time repeatedly instrumenting tracing code to each new version of the library and application to ensure accurate tracing functionality. Furthermore, the addition of tracing code to the libraries may result in version incompatibilities, and potentially introduce security risks, program bugs, or fatal errors.
To eliminate code staking for distributed tracing, non-invasive distributed tracing systems are explored to track end-to-end requests without any instrumentation into application source codes. 
Existing non-invasive tracing systems use aggregation algorithms based on commonly-used network metrics and metadata~\cite{deepflow, containiq, grpc, servicemesh}. 
However, these non-invasive implementations fail to reconstruct request causality and track end-to-end requests, making them impractical in the real-world scenario with highly concurrent requests. 

% 5. Challenges
To implement a practical non-intrusive tracing system that can track end-to-end requests of various applications and handle high concurrent requests, we will face the following challenges.

\begin{itemize}
    \item \textbf{Tracking complex end-to-end requests}: In a large-scale distributed system, multiple services have to coordinate and cooperate to serve a user request, resulting in complex request causality. It is difficult for non-invasive tracing systems to refactor request causality without instrumentation in source codes.
    \item \textbf{Compatible with heterogeneous high-level languages}: 
    Modern distributed systems are built in multiple high-level languages, such as Java, Python, and Golang. A non-invasive tracing system should be compatible with those commonly used high-level languages.
    \item \textbf{Supporting single-threaded and multi-threaded applications}: Applications serve requests in either single-threaded or multi-threaded manners. 
    Tracing systems should adapt to different implementations of applications and track accurate end-to-end requests for them.
    \item \textbf{Maintaining high performance}: Non-invasive tracing systems must maintain high performance even in large-scale distributed systems. The overhead of tracing should be minimized and not become a system bottleneck.
\end{itemize}

% 6. Solutions
To meet the challenges mentioned above, we propose a non-invasive tracing system, named Nahida, which is based on eBPF  (extended Berkeley Packet Filter)~\cite{corbet2014extending,ebpf222aaa}, a technology that allows user programs to be executed in the Linux kernel. Nahida first intercepts requests and responses with eBPF and injects custom trace contexts (e.g., trace ID) as tags into the messages. These contexts travel along with requests across the network from the downstream to the upstream service, and can be automatically injected into the subsequent requests without user awareness. 
Trace contexts are used to track end-to-end requests even in complex large-scale systems.
Nahida implements context injection and extraction in Linux kernel, which are decoupled from user application implementations. 
This allows Nahida to track end-to-end requests for applications implemented in multiple high-level languages, including Java, Python, Golang, etc.
Our evaluation results demonstrate that Nahida can track more than 92\% of end-to-end requests for single-threaded and multi-threaded microservices applications, even under high concurrency.
Moreover, Nahida has a little impact on service performance, such as a 1.55\%--2.1\% increase in service latency, an additional 1\%--3.78\% of CPU and memory consumption.

% 7. Contributions
Contributions of this paper can be summarized as follows.
\begin{itemize}
    \item We have implemented an in-kernel non-invasive distributed tracing system, Nahida, based on eBPF to successfully track end-to-end requests in large-scale distributed systems, regardless of application implementations in heterogeneous high-level languages and encapsulation in containers.
    \item We implement trace context propagation in the kernel to track end-to-end requests. With in-kernel context propagation, Nahida can adapt to single-threaded and multi-threaded application implementations.
    In contrast, existing invasive tracing systems fail in tracking end-to-end requests for multi-threaded applications with instrumentation in dependent third-party libraries, and so do the existing non-invasive tracing systems. 
    % experiment
    \item We conduct extensive experiments and successfully deploy Nahida in the real-world environment. Experiment results show that Nahida can achieve high tracing accuracy of more than 92\% with an insignificant overhead of 1.55\%--2.1\% latency increasing.
\end{itemize}

\begin{figure}[t]
    \centering
    \subfigure[An end-to-end request to the application.]{
        \includegraphics[width=0.9\linewidth]{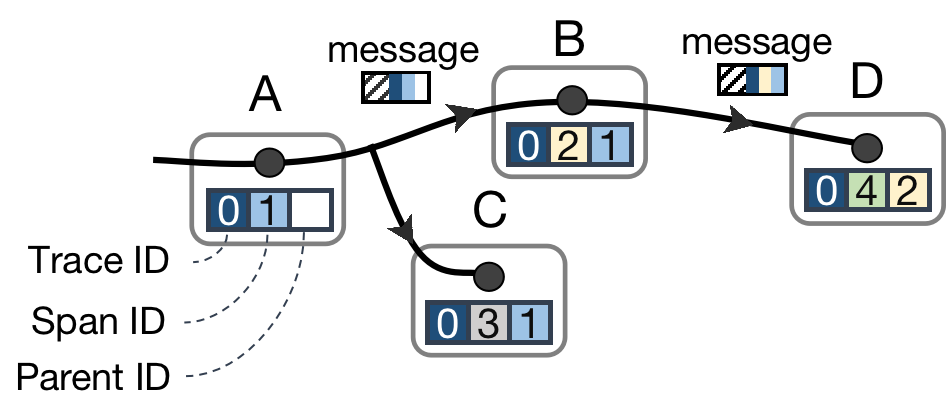}
        \label{fig:example_application}
    }
    \subfigure[A generated trace tree.]{
        \includegraphics[width=0.5\linewidth]{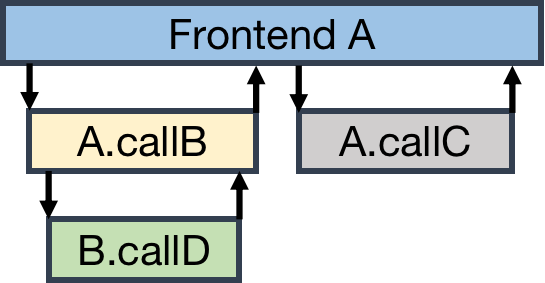}
        \label{fig:example_tracetree}
    }
    % \vspace{-0.1in}
    \caption{An example trace of a distributed application.}
    \label{fig:example}
\end{figure}

%-------------------------------------------------------------------------------

\section{Background}
\label{sec:background}
%-------------------------------------------------------------------------------

% \subsection{Terms}
\subsection{Distributed Tracing System}

Distributed tracing systems provide developers with the ability to track end-to-end requests which flow through a complex distributed system consisting of multiple microservices~\cite{pinpoint,xtrace,dapper}. By capturing and storing the request information, distributed tracing systems can offer insights into the performance of each service and its contribution to the overall system performance. Typically, information of an end-to-end request is displayed in a trace tree, which visualizes the path of a request as it travels through the system and the time each service takes to process the request. Analyzing trace trees can help developers identify bottlenecks and troubleshoot performance issues related to the request. As distributed systems and microservices architectures continue to gain in popularity, distributed tracing systems have become a crucial tool for managing and monitoring complex systems.

The following concepts are helpful in understanding  distributed tracing systems:
\begin{itemize}
    \item \texttt{Trace}: In a tracing system, a trace represents the complete request path for an end-to-end request. Each trace is identified by a unique \textit{trace ID}, which is used to correlate spans of the same trace across different services.
    \item \texttt{Span}: A span represents an operation in a trace, for example, service A calls service B. Each span is identified by a unique \textit{span ID} and is recorded with a start and end timestamp.
    \item \texttt{Parent Span}: A span can have a parent span, which indicates their causal relationship. The parent span is identified by a \textit{parent ID}, and helps to show the sequence of operations in an end-to-end request.
    \item \texttt{Root Span}: The root span is the first span in a trace and has no parent span. It represents the entry of the request.
    \item \texttt{Trace Context}: The trace context contains the trace ID, span ID, parent ID, and other relevant information. It is passed along with transmitting messages among services and enables associating two causal requests.
\end{itemize}

Figure \ref{fig:example} shows an example to illustrate how a distributed tracing system works to help with failure troubleshooting.
The distributed tracing system can collect trace contexts with the same trace ID from four services of an end-to-end request to an application, and then construct a trace tree with four spans as shown in Figure \ref{fig:example_tracetree}. 
When a failure in service D happens, traveling through the trace tree from the root span with ID 1 to its children span 2 and then span 4, the failure can be troubleshot as soon as possible by comparing the serving duration of each span.
This example highlights the importance of propagating trace contexts and how distributed tracing systems enable efficient troubleshooting in large-scale distributed systems.

\subsection{eBPF}

BPF (Berkeley Packet Filter)~\cite{mccanne1993bsd} was initially designed for packet filtering and later extended as eBPF (extended BPF)~\cite{corbet2014extending,ebpf222aaa} to enable user-defined event-triggered programs to be executed in the Linux kernel. eBPF provides a set of hooks in the network, where eBPF programs can be attached, enabling better network functionality~\cite{10.1145/3243157.3243161, jouet2017bpfabric, baidya2018ebpf, katran22222aab, xhonneux2018leveraging, choe2020ebpf, spright}.

The kernel programmability provided by eBPF makes it possible to implement a non-invasive distributed tracing system. 
Nahida can attach eBPF programs on system calls and sockets to track end-to-end requests without instrumenting applications or modifying the Linux kernel.
Moreover, the key-value datastore called eBPF Maps, defined by eBPF, helps our tracing system store data in the kernel. For example, Nahida mainly utilizes \textit{hash} of eBPF Maps, which acts as a hash table. 
Benefiting from the implemented eBPF wrappers, such as Bcc~\cite{zannoni2015new}  and libbpf~\cite{libbpf}, we can rapidly develop eBPF programs in high-level languages. libbpf is chosen to realize Nahida for its low overhead.

% %-------------------------------------------------------------------------------

\section{Motivations}
\label{sec:movitation}
\subsection{Existing Distributed Tracing Systems}

Distributed tracing systems track end-to-end requests to help developers and operators better understand request causality and benefits for root cause localization and bottleneck troubleshooting.
Distributed tracing systems are initially implemented in an invasive manner.
Existing invasive tracing systems~\cite{zipkin, jaeger, opentelemetry, opentracing, dapper} require instrumenting tracing codes into programming libraries or application codes in advance so that traces can be automatically  or manually generated without any user awareness. 
Although these existing invasive tracing systems can achieve high accuracy and enhance fault troubleshooting efficiency by code instrumentation, they still process the following distinct drawbacks.

\textbf{High maintenance cost. } 
Existing invasive tracing systems require code instrumentation to track end-to-end requests, so they have to provide libraries that implement end-to-end request tracing in multiple high-level languages for developers to use and update their instrumented libraries once the high-level programming libraries change and the same for the application source codes. The frequent updating of libraries and applications makes library maintenance time-consuming, especially when a new high-level language comes out.

\textbf{Low accuracy with library instrumentation.} 
Existing invasive tracing systems can track end-to-end requests and generate accurate traces by injecting trace contexts into messages when applications send requests to the upstream (see Figure \ref{fig:example}). However, since applications can serve requests in multiple threads, we observed that existing invasive tracing systems with third-party library instrumentation are unable to track accurate end-to-end requests as implemented in application source codes in such complex application implementation.

Due to the above shortcomings of existing invasive systems, developers try to implement a general non-invasive distributed tracing system regardless of high-level programming languages, to reduce the maintenance costs of tracing code instrumentation.
Existing non-invasive tracing systems~\cite{deepflow, containiq, grpc, servicemesh} apply aggregation algorithms to reconstruct request causality based on the collected metadata. 
Although the existing algorithm-based tracing systems require zero code instrumentation, they \textbf{can not guarantee the accuracy} of the reconstructed request causality. As we observed, Deepflow~\cite{deepflow} tracks none of the end-to-end requests under a certain workload or complex applications because of the limitations of the iterative times and policy of its aggregation algorithm (detailed in Section \ref{sec:exp_q2} and \ref{sec:exp_q3}).

\subsection{Goals of Nahida}
The popularity of distributed applications puts forward some higher requirements for distributed tracing systems based on the pros and cons of existing distributed tracing systems, which can be summarized as the following four goals.
\begin{itemize}
    \item Goal 1: Zero instrumentation. To reduce the maintenance costs on code instrumentation, Nahida is considered to be implemented without instrumentation into applications or libraries. 
    \item Goal 2: Support for high-level languages. Nahida should perform well on applications in modern commonly-used high-level languages, such as C, Python, Java, etc. 
    \item Goal 3: High Accuracy. Tracking wrong request causality does harm to root cause localization when services fail in a distributed system. Nahida should keep high accuracy on tracking requests, even if applications serve requests in multiple threads. 
    \item Goal 4: Low Overhead. Nahida should track requests at low overhead on resource consumption and service performance, without introducing extra bottlenecks.
\end{itemize}

\section{System Design}
\label{sec:design}
\subsection{Overview}

\begin{figure}
    \centering
    \includegraphics[width=1\linewidth]{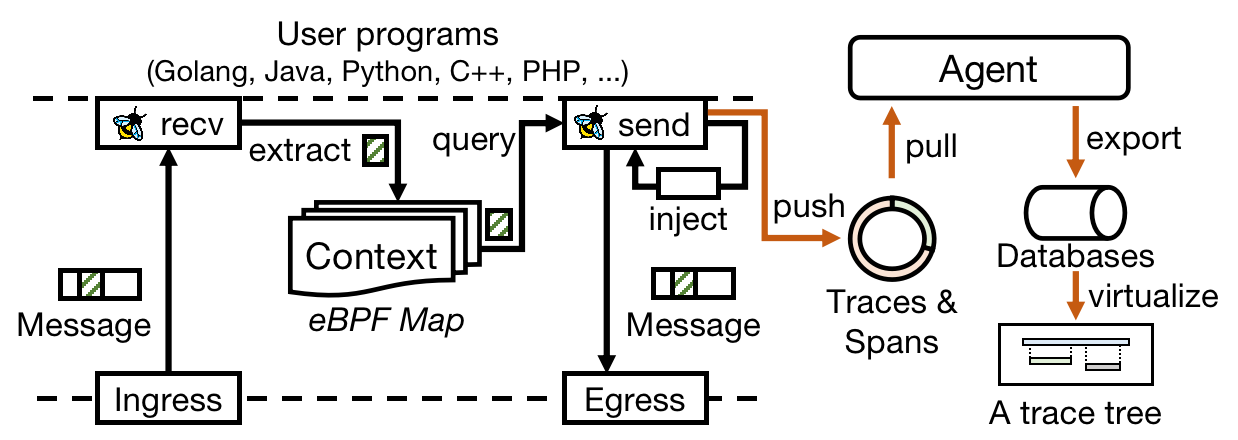}
    \caption{An overview of the workflow of Nahida. }
    \label{fig:overview}
\end{figure}

Nahida employs in-kernel eBPF to generate trace context and allows to propagate the generated contexts through user applications, regardless of their implementations in different high-level languages. To be compatible with existing distributed tracing systems, Nahida is designed to export the generated traces in the form of the popular standard, OpenTelemetry, making it possible to visualize traces in the existing user interface.

Nahida works as Figure \ref{fig:overview} shown. Nahida first attaches eBPF programs to the system functions on message transferring, such as \texttt{sendfile}, \texttt{sendmsg}, and \texttt{recvmsg}. The eBPF program on the receiving side parses and extracts trace contexts from the received message, and then stores it into an eBPF map. The eBPF program on the sending side queries the trace context and injects it into the message to be sent. If a request is finished, the eBPF programs will generate a trace with its spans and push it into a ringbuffer, an eBPF map. An agent is deployed on a host to make the generated traces accessible to users by pulling them from the ringbuffer, and exporting them to databases in OpenTelemetry standard. Users are able to access and visualize traces via their customized visualization consoles.

\subsection{Trace Generation}

To track end-to-end requests and generate traces, Nahida needs to generate a unique trace context for each operation in an end-to-end request, transmit the context alongside with messages, and propagate the context through user applications. More details in these procedures are shown as follows.

\subsubsection{Context Generation}
\label{sec:context_generation}

Inspired by OpenTelemetry and other existing distributed tracing systems, Nahida splits a request-response request into two separate operations, that is, the client requests the server and the server responds to the client. Each of the two operations can be viewed as a span in an end-to-end request. 
Nahida is designed to measure the time spent on these two operations of a request in the kernel. The time from the first sending of a request to the first receiving of a response is regarded as the client's time cost, while the time from the first receiving of a request to the first sending of a response is recorded as the server's time cost. 
For instance in Figure \ref{fig:example}, when a client sends a request to service A, service A takes the time from its first receiving of the request to the first sending of its response as its server time, and the time from the first sending of the request to service B to the first receiving of B's response as the client time to request B. 
We consider the span generated when the entry service receives and serves the entry request as the root span of a trace for an end-to-end request.
The other operations are all considered as children spans of that root span.

\begin{figure}
    \centering
    \includegraphics[width=1\linewidth]{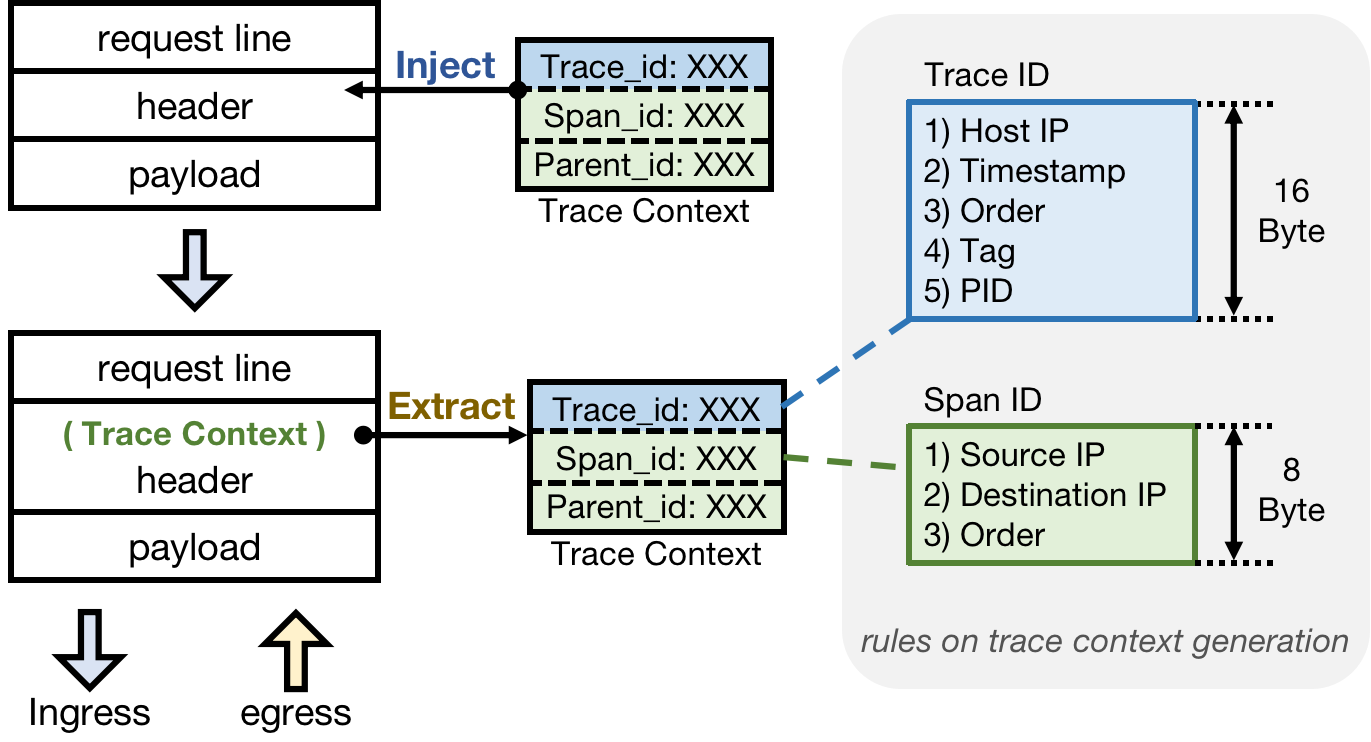}
    \caption{Rules for generating a trace ID and a span ID, when Nahida injects trace contexts into messages and extracts contexts from messages.}
    \label{fig:generation}
\end{figure}

To make end-to-end requests distinguishable, Nahida assigns a unique identifier to recognize each end-to-end request, that is trace ID. Similarly, operations in an end-to-end request are distinguished by span IDs. Rules for generating trace IDs and span IDs are displayed in Figure \ref{fig:generation}. 
Trace ID is a 16-byte value that includes the host IP address, the timestamp when the end-to-end request entries, a sequential order, a tag for debugging, and running process ID. Span ID is 8-byte, recording the IP addresses of the communicating client and server, as well as the span order in the whole trace.

\subsubsection{Context Transmission}

After generating a unique trace context for each operation in an end-to-end request, Nahida has to place those contexts alongside requests, so that Nahida can track end-to-end requests and establish accurate request causality.
The transmitting messages itself are chosen as the best way to carry trace contexts. By analyzing HyperText Transfer Protocol (HTTP)~\cite{fielding1999hypertext}, its key-value header is helpful in placing our generated trace contexts. As shown in Figure \ref{fig:generation}, when sending a message, Nahida injects trace contexts as our customized key-value pairs into the header of HTTP messages before the message is sent out of the host. When receiving a message alongside with our customized trace context, Nahida intercepts the message and then extracts the context from it.

Alongside messages, trace contexts can easily travel across the network, and can be successfully transferred between two services. In the meantime, since Nahida injects context into the HTTP header instead of the payload, the injection of contexts is free from the affection on the original messages, making the injected message be accurately processed by services.

\subsubsection{Context Propagation}
\label{sec:context_propagate}

The core of tracking an end-to-end request in a distributed tracing system is propagating the trace context through the service from one downstream request to its upstream request, to make accurate request causality(Goal 1).

The existing invasive tracing systems have implemented context propagation by instrumenting tracing codes in application source codes or programming libraries, but it is a challenge to propagate trace contexts in the kernel in a non-invasive way, without less knowledge of how a running application serves its user requests.
To meet this challenge and achieve the goals mentioned in Section \ref{sec:movitation}, we build up a non-invasive distributed tracing system. Based on the fact that an application is scheduled on a user process to execute by the operating system, we observe that the scheduling of an application on a user process is always changeless, where we just need to capture the user thread creation in the kernel to understand the execution of user threads as well as their parent-child relationships. % 
Since Nahida does not care about the detailed serving process of applications, it brings Nahida a possibility to \textbf{propagate trace context with the relationships of threads} in the kernel in a non-invasive way.

Applications serve requests mainly in the following two ways.
\begin{itemize}
    \item Single-threaded mode: Single-threaded applications serve a request in a single thread. If the serving thread has to request its upstream for services, it will send the request and wait for the response in the same thread.
    \item Multi-threaded mode: Multi-threaded applications always serve a request by multiple collaborating threads. When the serving thread asks its upstream for services, the application chooses to create or schedules its idle thread to request the upstream and then keeps serving the user request with no wait.
\end{itemize}

Based on these two implementations of applications, we discuss how Nahida propagates trace contexts.

\begin{figure}
    \centering
    \subfigure[A single-threaded application.]{
        \includegraphics[width=0.5\linewidth]{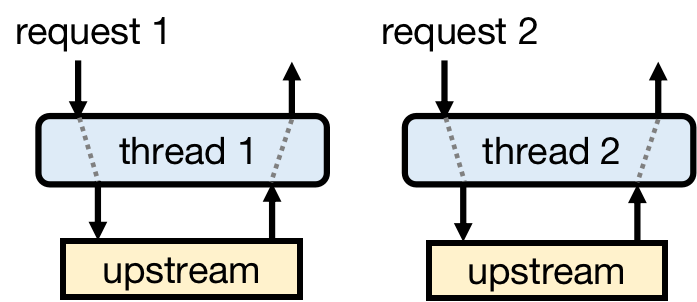}
        \label{fig:single_thread}
    }
    \subfigure[A simple multi-threaded applications.]{
        \includegraphics[width=0.6\linewidth]{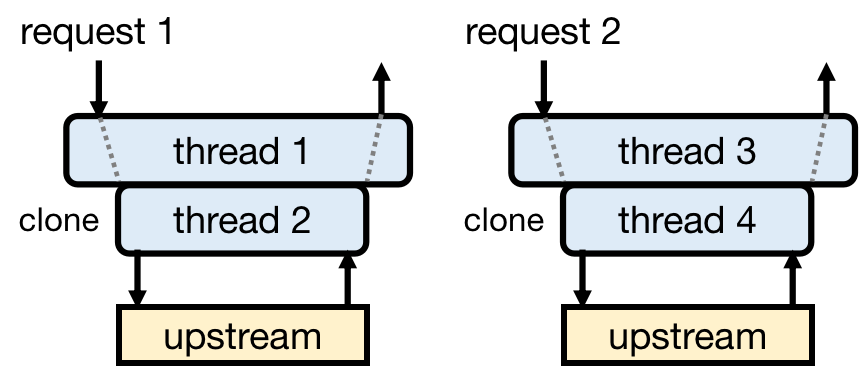}
        \label{fig:asyn_mode}
    }
    \caption{Implementation of single-threaded and multi-threaded applications.}
    \label{fig:implementation_application}
\end{figure}

For \textbf{single-threaded applications}, it is easy for Nahida to propagate trace contexts through applications with the unique identification of a thread (thread ID). Since a thread serves a request synchronously (see Figure \ref{fig:single_thread}), Nahida can first store the extracted trace context with thread ID as key when receiving a request from downstream, and then queries that trace context for the same thread ID and injects that context into the message which will be sent to the upstream service. As a result, trace contexts can be propagated through the end-to-end requests.

For \textbf{multi-threaded applications}, they parallelize requesting the upstream service and keep serving the received user request themselves so as to improve their performance. 
In a multi-threaded application, once a thread serves a request from its downstream, it forks and clones a new thread to request for its upstream (see Figure \ref{fig:asyn_mode}). By monitoring the thread creation, Nahida can construct the parent-child relationships among threads. Based on these relationships, Nahida can implement context propagation. Nahida first stores the received trace context with the parent thread ID as the key. When the child thread sends a message to the upstream, Nahida finds the parent of that child thread, then figures out the trace context of that parent thread and injects it into the sending message.

\subsection{Trace Collection and Visualization}

After generating trace contexts and propagating them across the network and through applications, Nahida can generate a unique trace for each end-to-end request. Traces with their spans are pushed into a ringbuffer, an eBPF map, and then pulled by an agent deployed in the host. For long-term storage, the agent exports those traces to local or external databases.

Traces in the databases can be accessed and visualized in the user interface to better help developers understand and debug end-to-end requests. To be compatible with the existing visualization user interface, our agent transforms the generated traces into the standard format of OpenTelemetry before exporting them into databases.

%-------------------------------------------------------------------------------

\section{Implementation}

\textbf{Context generation}: Nahida generates 8-byte span IDs and 16-byte trace IDs according to the rules of Section \ref{sec:context_generation}. The time when Nahida generates a trace context is the time when a service sends a request to its upstream and a service receives a request. By parsing the content of messages, Nahida can recognize the HTTP message by matching the HTTP format and then generate trace context, without affecting other messages not in HTTP. 
In addition, thanks for the service governance provided by Kubernetes~\cite{bernstein2014containers}, each service and its instances have their instance IP addresses.
By filtering, Nahida can further reduce the time overhead on message parsing and request tracing for non-service requests, thus more effectively serving the end-to-end request tracing for microservice systems.

\textbf{Context transmission}: The generated trace context needs to be injected into the sending message and extracted when received. Benefiting from the eBPF helper function, \texttt{bpf\_msg\_push\_data}, Nahida can intercept the sending message and inject our customized context into it at a specific location without any user awareness. Nahida utilizes a \texttt{sk\_msg} program to split a message into two fragments in a scatter list, then injects our customized context in the first fragment after the request line, where the first key-value data in the HTTP header is placed. The request line can be recognized by the hexadecimal ending symbol \texttt{0x0d0a} of each line in an HTTP message. By doing so, Nahida can reduce the number of operating instructions required for localization finding, as the eBPF program has an upper limit on the number of instructions due to its security considerations.
After context injection, these two fragments are merged back as a new message to be sent. When received the request alongside the unique context, our \texttt{kprobe} and \texttt{kretprobe} programs hooked on the kernel function \texttt{sock\_recvmsg} help to obtain the content of the  received message and extract our injected trace context.

\textbf{Context propagation}: As analyzed in Section \ref{sec:context_propagate}, to propagate trace contexts through applications, Nahida must understand parent-child relationships of threads. 

\begin{figure}
    \centering
    \includegraphics[width=0.7\linewidth]{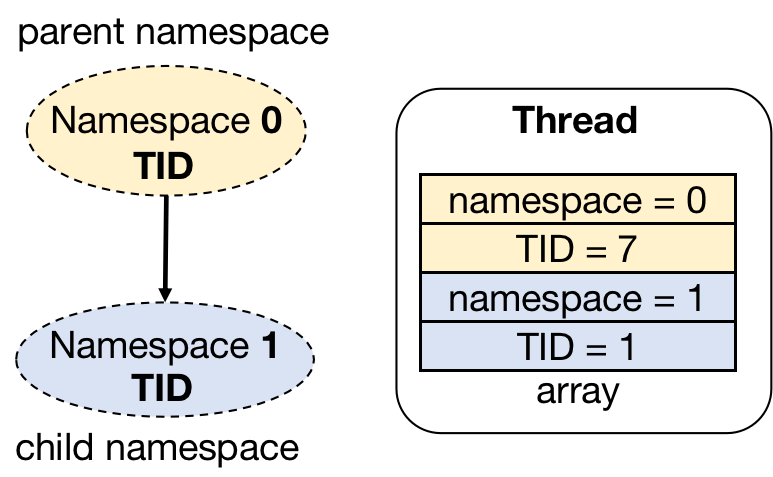}
    \caption{Thread ID assignment in Linux. Each thread ID in the child namespace has a unique ID corresponding to its parent namespace.}
    \label{fig:pid_ns}
\end{figure}

In microservices, applications are deployed in containers to achieve fast deployment.
However, resource isolation among containers in namespaces makes it a challenge to build up parent-child relationships of threads in the kernel. 
Threads of a containerized application execute in their container's namespace, and are assigned unique IDs in that namespace. In addition, threads in the children namespaces should also own unique IDs in its parent namespace. 
The IDs of a thread are different from namespaces, where Nahida executes in the root namespace while applications run in its children namespace. When monitoring the system function \texttt{fork} and \texttt{clone}, Nahida obtains the IDs of children threads in the children namespace and the IDs of  parent threads in the root namespace, which introduces inconsistency of thread IDs and makes inaccurate parent-child relationships among threads. 
To meet this challenge, we try to figure out how Nahida obtains the thread IDs in different namespaces. Figure \ref{fig:pid_ns} shows the data structure of a thread in Linux, where a thread is regarded as a process to execute. A thread records its namespace in the variable called \texttt{level} and the ID of each namespace in an array named \texttt{number}. When a new thread is forked, it is allocated new IDs in all its located namespaces by calling the system function \texttt{alloc\_pid}. The \texttt{kprobe} program on that function can help Nahida to obtain the structure of a thread and parse its IDs in all its namespace, thus Nahida is able to keep the consistency when establishing the parent-child relationships of threads.

With the collected information, to realize context propagation, Nahida stores the extracted context with thread ID as a key. When sending requests to upstreams, Nahida finds the context to be injected. 
When a thread sends a message to upstreams, Nahida first looks for the trace context with running thread ID. If the running thread does not have its own trace context, Nahida continues to look for the trace context of its parent thread. After finding the trace context, Nahida will inject the child context of that trace context into the message to upstreams, thereby propagating trace contexts through applications.

\textbf{Traces exportation.}
Our eBPF programs export traces and spans when a client receives its response and when a server sends the response. 
To meet the requirement for adapting to the existing tracing user interface, the agent on each host tries to read our generated traces and spans from the ringbuffer, and then export them to the external databases, with the standard of OpenTelemetry.

\section{Evaluation}
\label{sec:evaluation}
To evaluate the performance of Nahida, we conduct a series of experiments and mainly answer the following questions.

\textbf{Q1}: How Nahida work with applications implemented in multiple high-level languages?

\textbf{Q2}: How well Nahida can perform in the exampled microservice applications?

\textbf{Q3}: How stably does our tracing system perform under different concurrency?

\textbf{Q4}: How much is the extra overhead  of Nahida on resource consumption and service performance?

\subsection{Experiment Settings}
\subsubsection{Environment}
We have set up our test environment on a Kubernetes~\cite{bernstein2014containers} platform with 5 physical machines, each having 16GB RAM and eight 4-core 2.50GHz Intel Xeon Platinum 8269 CPUs. The machines run on Centos-8-based OS and have Linux kernel version 5.10.134. To compile the eBPF programs, we have opted for \textit{clang} and \textit{LLVM} version 13. We have used Fortio~\cite{fortio} to generate user-defined loads to our benchmarks, while the \textit{sar} command has been utilized to obtain the CPU and memory usage of each machine.

\subsubsection{Benchmark}

For our experiments, we have deployed two open-source microservice application benchmarks as our testbed. The first benchmark is Bookinfo~\cite{bookinfo}, which is an online bookstore serving in microservices. Bookinfo is written in various programming languages, including Python, and Java. The second benchmark is Trainticket~\cite{trainticket}, which is a large-scale distributed train ticket booking system consisting of 41 microservices, written in Java. 
Each service of these two benchmarks owns one replica. Both Bookinfo and Trainticket are single-threaded, and thus they are regarded as our benchmarks on single-threaded applications.
Since these two benchmarks do not provide multi-threaded examples, we refactored the source code of the frontend service of Bookinfo and Trainticket to build up multi-threaded applications, which can be regarded as our benchmarks on multi-threaded applications.

In addition, we conducted both single-threaded and multi-threaded benchmarks in multiple high-level languages, including PHP, Java, Python, and Golang to test how Nahida performs on them. Each example benchmark consists of three services, where the frontend service asks the middle service for content and the middle requests the backend.

\subsubsection{Baselines}

Jaeger with OpenTracing~\cite{jaeger, opentracing} is a commonly used approach for tracking end-to-end requests in a distributed system. Jaeger stands for a typical invasive implementation of a distributed tracing system, which adheres to the OpenTelemetry/OpenTracing standard and instruments tracing codes directly into service source codes and can be considered as the most accurate baseline.
Deepflow~\cite{deepflow} is state-of-the-art non-invasive approaches for tracking requests, which implements an aggregation algorithm based on its collected metadata. We compare Nahida with Deepflow to show our performance.

\subsubsection{Evaluation Metrics}
To better evaluate the effectiveness of Nahida, we use the precision as our metric. Precision reflects the similarity between the request traces generated by different tracing systems and the actual request invocation relationship. The higher the precision, the more effective the tracing system.
To measure the impact of Nahida on service performance, we use two metrics: request latency and query-per-second (QPS). Request latency is used to show the impact of Nahida on service response time. QPS is used to measure the throughput of the service. In addition, we also measure the consumption of CPU and memory to help understand the resource consumption of Nahida on the hosts. By using these metrics, we can analyze the overhead of Nahida.

\subsection{(Q1) Multi-language Supporting}

\begin{figure}
    \centering
    \includegraphics[width=1\linewidth]{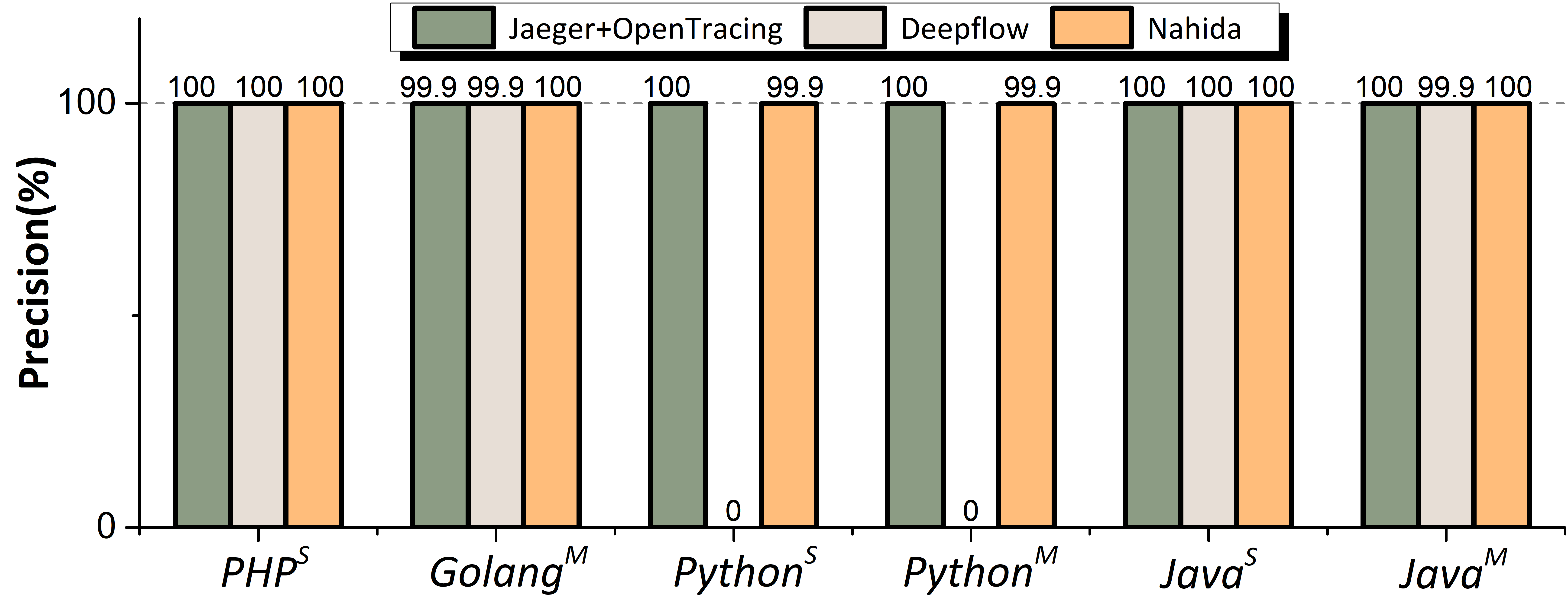}
    \caption{Effectiveness comparison between Nahida, Deepflow, and Jaeger with OpenTracing when they track requests served by multiple high-level languages.}
    \label{fig:exp1}
\end{figure}

To test the performance of Nahida on supporting multi-language application tracing, we used Fortio to generate a total of ten thousand requests continuously and sequentially for our constructed single-threaded and multi-threaded benchmarks in multiple high-level languages. We collect the evaluation metrics and draw Figure \ref{fig:exp1}, where ``S'' represents the service responses in single-threaded mode, while ``M'' represents replying in multi-threaded mode. We test PHP just in single-threaded mode due to its widespread use. 

As shown in Figure \ref{fig:exp1}, Jaeger can always achieve perfect performance on tracing end-to-end requests by instrumenting tracing codes into applications in multiple high-level languages.
Deepflow also supports multi-language applications. With the monitored data on network communication, Deepflow analyzes causality between requests, regardless of the implementation of applications. As we can see, Deepflow can aggregate almost all request causality for PHP, Golang, and Java applications, but it does bad in tracking end-to-end requests for Python applications, no matter if the application is implemented in the single-threaded mode or multi-threaded mode. By recognizing the trace log collected from Deepflow, we find out that requests can not be correctly associated with the value of a tag syscall\_trace\_id, which is often zero, resulting in a failure to restore the complete end-to-end trace. 

Nahida can track nearly 100\% of end-to-end requests of multi-language applications, even those applications serve requests in multi-threaded mode. Monitoring parent-child relationships of threads, Nahida can recognize in the kernel which thread that a request belongs to, thus propagating trace contexts through applications from one thread to another thread, without detailed invasive information of applications. The result shows that Nahida performs well in tracking the end-to-end requests in a non-invasive way.

\subsection{(Q2) Microservices Examples}
\label{sec:exp_q2}

\begin{figure}
    \centering
    \includegraphics[width=1\linewidth]{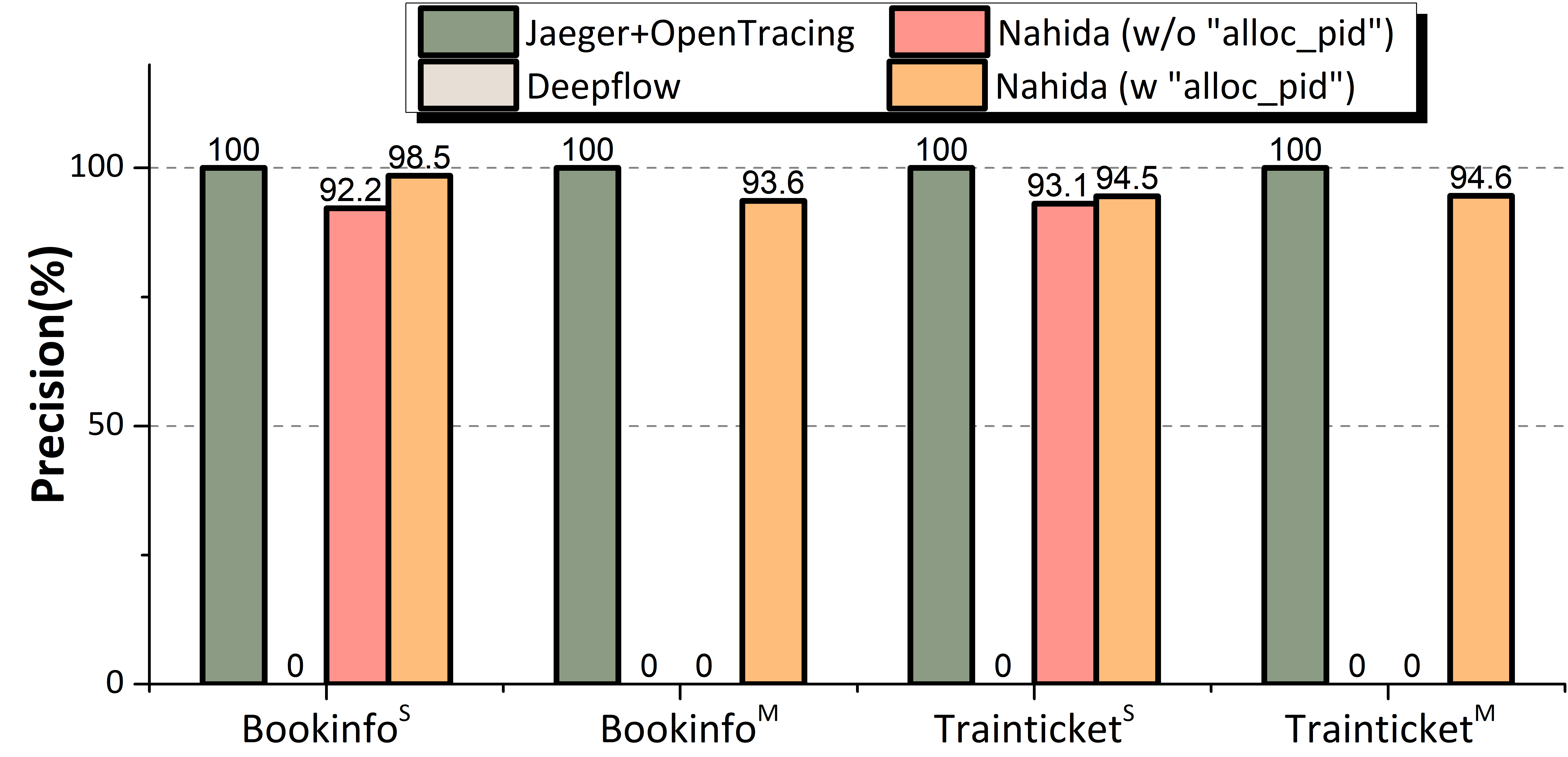}
    \caption{Effectiveness comparison between Nahida, Deepflow, and Jaeger with OpenTracing when they track requests for single-threaded and multi-threaded Bookinfo and Trainticket.}
    \label{fig:exp2}
\end{figure}

To evaluate the performance of Nahida on tracking end-to-end requests of containerized microservices, we choose Bookinfo and Trainticket as our benchmarks. Fortio is set to generate a total of ten thousand requests continuously and sequentially for the single-threaded and multi-threaded Bookinfo and Trainticket. The comparison result on accuracy is shown in figure \ref{fig:exp2}. 

Jaeger still keeps tracking end-to-end requests perfectly with instrumentation into application codes. 
Deepflow fails to reconstruct accurate causality between requests, since the tags named \texttt{syscall\_trace\_id} of trace logs are always zero which shows no causality relationships among requests and makes a parent request lose its child request and thus fails to generate an end-to-end trace. 
Meanwhile, the aggregation algorithm of Deepflow has a limit of the iterative times. When the number of service calls involved in an end-to-end request is too many, all the involved service calls cannot be traversed under the limited iterative times. For example, for Travel service requests with 97 spans under Trainticket, Deepflow could not restore the end-to-end requests.

Nahida performs better than Deepflow even tracking end-to-end requests of containerized microservices. Since a container works as a process in the host, Nahida can easily find the parent-child relationship among its threads and propagate trace context through the application, thus achieving accuracy on tracking requests of at least 93.6\%, even tracking for a total of 97 spans for an end-to-end request to Trainticket. The inaccurate and incomplete traces are generated when the Detail service or the Reviews service does not generate their spans normally. By observing the logs from eBPF, we find that after the service receives a request and records its trace context, the eBPF program is not triggered when sending the response, resulting in no span generated on the server side, thus affecting the accuracy of the trace generation.

From Figure \ref{fig:exp2}, We can observe that the eBPF program hooked on function \texttt{alloc\_pid} helps Nahida well to build a parent-child relationship for containerized service threads on the host, regardless of the resource isolation of namespaces. Without monitoring the function \texttt{alloc\_pid}, Nahida tracks end-to-end requests well of the single-threaded Bookinfo and Trainticket application, since a thread processes a request individually. However, because of the resource isolation of namespaces, Nahida can not find the parent-child relationships between two threads in different namespaces, so Nahida monitoring the function \texttt{sys\_clone} instead of \texttt{alloc\_pid} fails to track end-to-end requests of multi-threaded Bookinfo and Trainticket.

\subsection{(Q3) Concurrency Impact}
\label{sec:exp_q3}

\begin{figure}
    % \centering
    \includegraphics[width=1\linewidth]{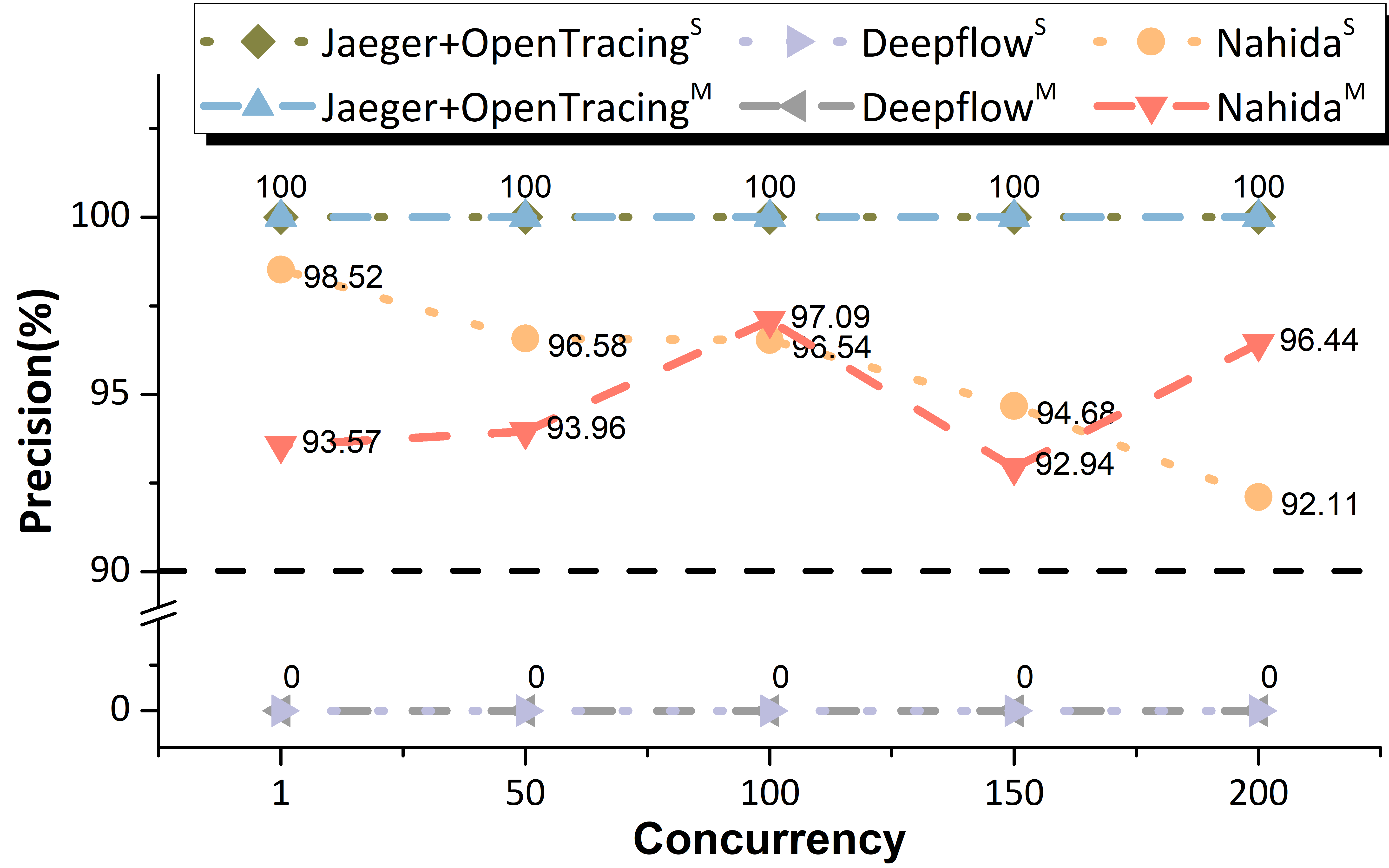}
    % \vspace{-0.1in}
    \caption{Effectiveness comparison between Nahida, Deepflow, and Jaeger with OpenTracing when they track concurrent requests for single-threaded Bookinfo.}
    \label{fig:exp3}
\end{figure}

To evaluate the performance of Nahida under different concurrency, we selected Bookinfo as our benchmark since it can process up to two hundred concurrent requests simultaneously, whereas Trainticket can only process less than one hundred requests at the same time. Fortio is configured to generate a total of ten thousand requests for both single-threaded and multi-threaded Bookinfo, with 1, 50, 100, 150, and 200 concurrent users, respectively. Compared results with Jaeger and Deepflow are presented in Figure \ref{fig:exp3}.

Jaeger keeps tracking all the end-to-end requests of either single-threaded or multi-threaded applications under different concurrent user workloads. With explicit instrumentation in application source codes, Jaeger can easily distinguish which downstream request caused each concurrent upstream request, thus tracking the exact causal relationship between requests.

In the case of concurrent user requests, Deepflow can aggregate part of the Bookinfo application's end-to-end requests, but it cannot support aggregating the complete end-to-end requests with a metadata-based algorithm, so the accuracy of its reconstructed request causality is almost zero. For example, the Productpage service needs to request the Detail service and the Reviews service, respectively, to make a response to a user request. Deepflow treats these two requests as two separate traces rather than a trace for an end-to-end request. 
Observed from trace logs collected by Deepflow, Deepflow utilizes a tag named \texttt{syscall\_trace\_id} to find out the causality relationships between two trace logs, but it fails to assign accurate syscall\_trace\_id to associate the request from Productpage to Details with the request from Productpage to Reviews, which are likely to be zero. In addition, Deepflow applies the policy that the multiple requests or responses in the same transmission direction share the same trace ID, so that it confuses the causal relationship between many requests. 

% ours
Nahida can achieve accuracy at more than 92\% when tracking concurrent end-to-end requests for single-threaded and multi-threaded applications. By propagating trace context with parent-child relationships of threads, Nahida can perform well on request tracking, stably. The reason why Nahida can not achieve 100\% accuracy is that the eBPF program hooked on the sending and receiving system functions can not perform properly due to high concurrency. As a result, Nahida fails to generate some of the spans.

\subsection{(Q4) Overhead}

\begin{table}
  \centering
  \fontsize{8}{11}\selectfont
  \caption{Overhead Comparison between the normal (Nahida-) and Nahida (Nahida+)}
  \vspace{0.05in}
  \label{tab:overhead}
  \begin{tabular}{ccccc}
    \toprule
    \toprule
    \textbf{Metric} & \textbf{Concurrency} & \textbf{Nahida-} & \textbf{Nahida+} & \textbf{Overhead ($\Delta$)}\\
    \midrule
    \multirow{2}{*}{Latency} & 100 & 683.85 & 694.67 & 10.82 (1.58\%) \\
    \cmidrule{2-5}
    ~ & 200 & 1374.76 & 1403.73 & 28.97 (2.1\%) \\
    \midrule
    \multirow{2}{*}{QPS} & 100 & 146.17 & 143.9 & 2.27 (1.55\%) \\
    \cmidrule{2-5}
    ~ & 200 & 145.33 & 142.33 & 3.0 (2.06\%) \\
    \midrule
    \multirow{2}{*}{CPU} & 100 & 10.65 & 10.76 & 0.11 (1.0\%)  \\
    \cmidrule{2-5}
    ~ & 200 & 10.49 & 10.83 & 0.34 (3.3\%)  \\
    \midrule
    \multirow{2}{*}{Memory} & 100 & 72.27 & 74.32 & 2.05 (2.84\%)  \\
    \cmidrule{2-5}
    ~ & 200 & 72.12 & 74.84 & 2.73 (3.78\%)  \\
    \bottomrule
    \bottomrule
\end{tabular}
\end{table}

We generated workloads with 100 and 200 concurrent users using Fortio to evaluate the overhead of Nahida when tracking requests for the single-threaded Bookinfo. The experiments run for three minutes and were repeated three times on five physical machines. The average results are presented in Table \ref{tab:overhead}.

First, we examine the impact of Nahida on the service performance. As shown in Table \ref{tab:overhead}, Nahida increases the request latency by about 10.82--28.97 ms (1.58\%--2.1\%) compared to the original, and reduces the QPS of the service by about 2.27--3.0 (1.55\%--2.06\%), even under high concurrency, such as two hundred. The additional latency comes from our customized eBPF programs when Nahida intercepts and parses the sending and receiving messages in the Linux kernel.

Regarding resource consumption, Nahida has increased the average CPU usage of machines by 0.11\%--0.34\% (1.0\%--3.3\% relative to the original).
Our eBPF programs executing in the Linux kernel cause extra CPU consumption. In the meantime, Nahida needs 2.05\%--2.73\% (about 336MB--447MB) more of the memory to temporarily store part of the messages, the generated trace context and other information.

As demonstrated in Table \ref{tab:overhead}, Nahida has a stable impact on service performance and resource consumption, regardless of the number of concurrent requests served by the service. Overall, compared to the original, Nahida has a 1.55\%--2.1\% loss in service performance and requires an additional 1.0\%--3.78\% CPU and memory consumption.

\subsection{Discussion}
\label{sec:discussion}
We implement a non-invasive distributed tracing system, Nahida, for applications serving the HTTP message. There exist some limitations in our implementation, which lead to our future work.
First, messages in protocols other than HTTP, such as HTTPS and gRPC, are encrypted, which makes it difficult for Nahida to inject trace contexts into those messages in the kernel, due to their unstructured data. In this case, we can inject the trace context at the front of a request instead of in the header to propagate trace contexts between services in our future work. This approach would require the receiver to deploy an in-kernel eBPF program to resolve and remove the context from the message without affecting the upstream service serving request messages.
Second, eBPF programs can only be deployed by privileged users. But Linux has the plan to make them available to unprivileged users~\cite{Unprivileged_eBPF}, and there exists some work on unleashing unprivileged eBPF programs~\cite{10.1145/3609021.3609301}. The requirement to root privilege of eBPF programs may be alleviated in the future.
Third, the mounting of eBPF programs to some user functions requires ZeroTracer to obtain the addresses of the user functions in different third-party libraries.
Fourth, ZeroTracer requires a Linux kernel version of at least 5.2~\cite{kernel_version} to execute.
Fifth, ZeroTracer is compatible with the existing OpenTelemetry and does not affect request tracing of OpenTelemetry.

\section{Related Work}
\label{sec:related}

Invasive tracing systems require developers to instrument tracing codes into their service codes to generate traces. Zipkin~\cite{zipkin}, Jaeger~\cite{jaeger}, SkyWalking~\cite{skywalking}, OpenTracing\cite{opentracing}, OpenTelemetry~\cite{opentelemetry}, and Dapper~\cite{dapper} are some typical examples of invasive tracing systems. In these tracing systems, applications or programming libraries must be instrumented either in the requester's codes or in the requestee's codes. 
Castanheira \textit{et al.}~\cite{flexible} proposed a flexible distributed tracing workflow based on instrumented annotations and preemptive tracing for debugging.
JCallGraph~\cite{jcallgraph} realizes a tracing system for microservice applications by instrumenting the cloud platform middlewares.
However, as applications and libraries are updated frequently, developers have to repeatedly instrument the updated codes, which consumes significant time. Instrumenting the programming libraries can achieve user transparency, but applications need to be executed in the instrumented environment. On the other hand, instrumented applications can be executed in any environment, but require explicit instrumentation by the developer.

Existing non-invasive tracing systems generate traces of requests without any instrumentation into applications and programming libraries. 
Inkle~\cite{grpc} tracks gRPC requests for microservice applications on Kubernetes. Cha \textit{et al.}~\cite{servicemesh} constructed the request traces based on the information collected by the service proxies in the service mesh. Both approaches can only generate single-hop traces without instrumenting the source codes.
DeepFlow~\cite{deepflow}, and ContainIQ~\cite{containiq} mount eBPF programs at some of the Linux system calls and user-defined functions to collect commonly-used metrics and generate traces by an aggregation algorithm or metadata-based algorithm. However, both approaches may suffer from a loss of accuracy.

%-------------------------------------------------------------------------------

%-------------------------------------------------------------------------------

\section{Conclusion}
\label{sec:conclusion}
To track requests in a large-scale distributed system in a non-invasive way, we propose Nahida, based on eBPF, regardless of high-level programming language and application implementations. Our evaluations demonstrate that Nahida can track completely over 92\% of requests, maintaining stable tracing performance with increased concurrency. Notably, Nahida can fully track requests served by multi-threaded applications, a capability that existing invasive tracing systems that instrument tracing codes in programming libraries lack. Furthermore, our tests indicate that Nahida has little impact on service performance, with only a 1.55\%--2.1\% increase in service latency. In terms of resource consumption, Nahida consumes an additional 1.0\%--3.78\% of CPU and memory resources, compared to the normal system.

% \bibliographystyle{plain}
% \bibliography{reference.bib}

%%%%%%%%%%%%%%%%%%%%%%%%%%%%%%%%%%%%%%%%%%%%%%%%%%%%%%%%%%%%%%%%%%%%%%%%%%%%%%%%
\end{document}